\documentclass[sigconf, 10pt, nonacm]{acmart}
\usepackage{amsmath}

\usepackage{graphicx}
\usepackage{subcaption}
\usepackage{stfloats}
\usepackage{tabularx}
\usepackage{booktabs}

\usepackage{enumitem}

\newlist{fitemize}{itemize}{1}
\setlist[fitemize]{
 leftmargin=0em,      
  labelwidth=1.5em,      
  labelsep=0em,          
  itemindent=1.5em,     
  label=\textbullet,     
  align=left,            
  listparindent=0pt
}

\newcommand{\swap}{\sigma}
\newcommand{\rateup}{{r_{+}}}
\newcommand{\ratedn}{{r_{-}}}
\newcommand{\distill}{{\mathcal D}}
\newcommand{\decohererate}{{\mathcal L}}
\newcommand{\bcount}{{\mathcal C}}
\newcommand{\iswap}[3]{{#2$\leftarrow$#1$\rightarrow$#3}}
\newcommand{\share}[1]{[#1]}
\newcommand{\setofnodes}{{\mathcal N}}
\newcommand{\numnodes}{|\setofnodes|}

\title{\texorpdfstring{Path-Oblivious Entanglement Swapping for the Quantum Internet \newline {\large July 9, 2025}}{My title baby}}

\author{Vincent Mutolo}
\affiliation{%
  \institution{Columbia University}%
  \city{New York, NY}
  \country{USA}%
}

\author{Rhea Parekh}
\affiliation{%
  \institution{Columbia University}%
  \city{New York, NY}
  \country{USA}%
}
\author{Dan Rubenstein}
\affiliation{%
  \institution{Columbia University}%
  \city{New York, NY}
  \country{USA}%
}

\begin{document}

\begin{abstract}
 Proposed Bell pair swapping protocols, an essential component of the Quantum Internet, are planned-path: specific, structured, routing paths are reserved prior to the execution of the swapping process.  This makes sense when one assumes the state used in the swapping process is expensive, fragile, and unstable.  However, lessons from classical networking have shown that while reservations seem promising in concept, flexible, reservation-light or free approaches often outperform their more restrictive counterparts in well-provisioned networks.  In this paper, we propose that a path-oblivious approach is more amenable to supporting swapping as quantum state evolves into a cheaper, more robust form.  We formulate the swapping process as a linear program and present and evaluate a fairly naive baseline swapping protocol that tries to balance Bell pairs throughout the network.  Preliminary results show that while naive balancing leaves room for improvement, investigating path-oblivious swapping is a promising direction.
 \end{abstract}

\maketitle

\section{Introduction}

\begin{figure}[ht] 
    \centering
    \begin{subfigure}[b]{0.48\columnwidth}
        \includegraphics[width=\linewidth]{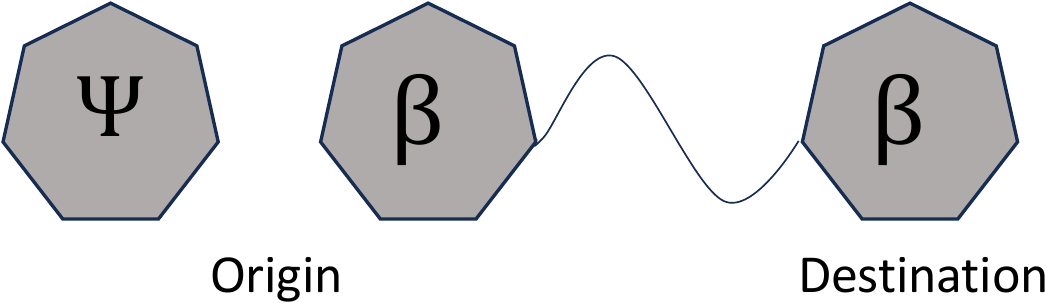}
        \caption{Initial configuration}
        \label{fig:tel-a}
    \end{subfigure}
    \hfill
    \begin{subfigure}[b]{0.48\columnwidth}
        \includegraphics[width=\linewidth]{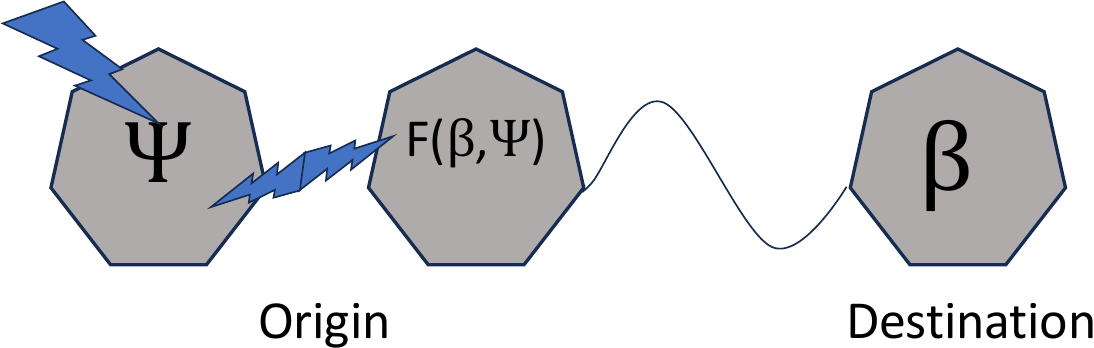}
        \caption{Origin Local Operations}
        \label{fig:tel-b}
    \end{subfigure}
    
    \vspace{0.5em} 
    
    \begin{subfigure}[b]{0.48\columnwidth}
        \includegraphics[width=\linewidth]{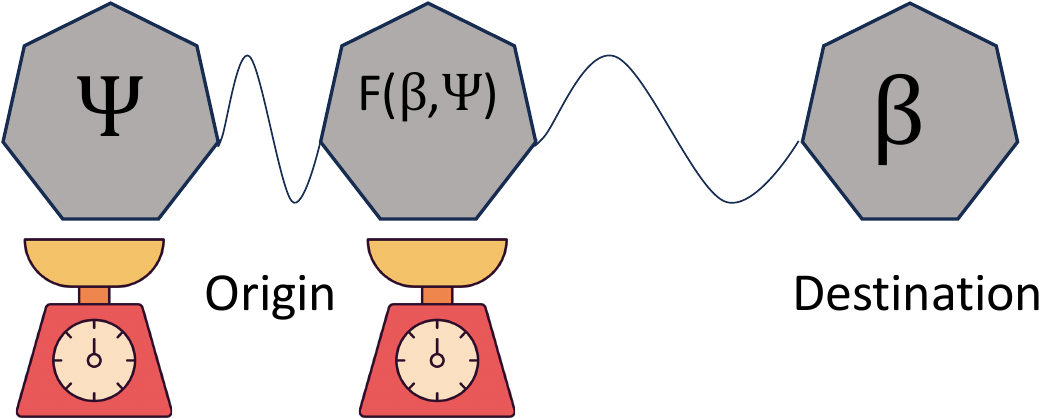}
        \caption{Origin measurement}
        \label{fig:tel-c}
    \end{subfigure}
    \hfill
    \begin{subfigure}[b]{0.48\columnwidth}
        \includegraphics[width=\linewidth]{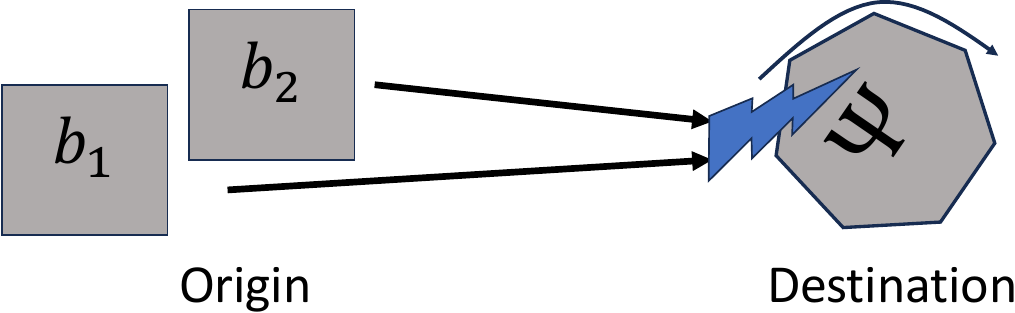}
        \caption{Classical bit transmission and Destination Repair}
        \label{fig:tel-d}
    \end{subfigure}
    \vspace{-12pt} 
    \caption{Teleportation process}
    \Description{A visual representation of the teleportation process in 4 steps}
    \label{fig:teleportation}
\end{figure}

The Quantum Internet is expected to make extensive use of {\em teleportation}~\cite{bennett1993teleporting}: the transfer of an aritrary quantum state (represented by a qubit) from one physical location (origin node) to another possibly distant location (destination node) via a sequence of local operations and transmission of only 2 bits of classical (non-quantum) information to complete the process, as summarized in Fig.~\ref{fig:teleportation}.  Aside from the qubit holding the state to be teleported, the process requires a Bell pair: two qubits that may be separated at an arbitrarily large distance, but remain {\em entangled}, meaning that the measurement of one of the qubits will affect the outcome of the subsequent measurement of the other.  The local operations include sequences of gate transformations (depicted as blue bolts) and measurement of qubits that collapses the qubit's state into a classical bit (1 or 0).

\begin{figure}[h] 
    \centering
    \begin{subfigure}[b]{0.48\columnwidth}
        \includegraphics[width=\linewidth]{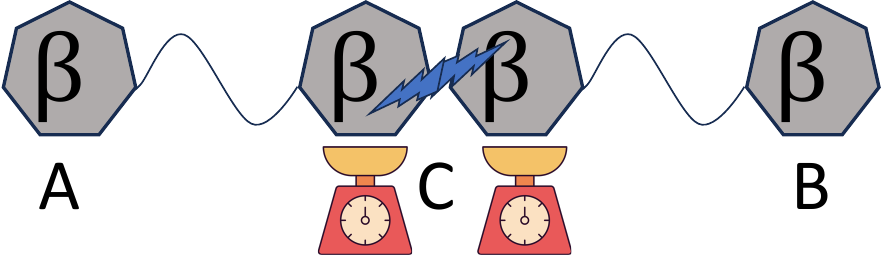}
        \caption{Before swap}
        \label{fig:swap-1}
    \end{subfigure}
    \hfill
    \begin{subfigure}[b]{0.48\columnwidth}
        \includegraphics[width=\linewidth]{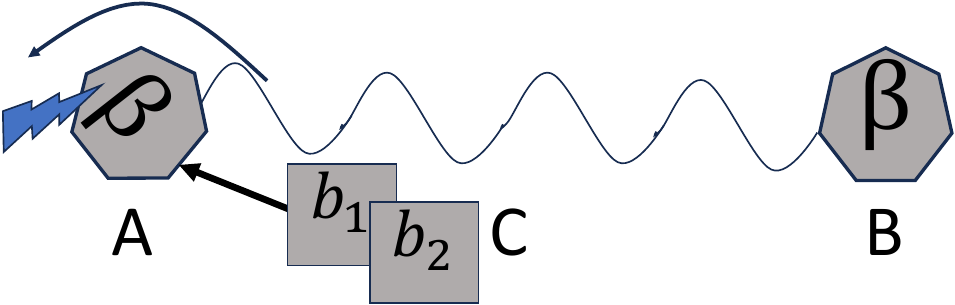}
        \caption{After swap}
        \label{fig:swap-2}
    \end{subfigure}
    \vspace{-12pt} 
    \caption{Swap operation}
    \Description{A pictorial demonstration of a swap operation}
    \label{fig:swap}
\end{figure}

Teleportation clearly depends on the origin and destination sharing a Bell pair.  While some pairs of nodes can directly {\em generate} these entangled pairs,  others, often due to geographical distance, cannot, and it is challenging to physically move the qubits without corrupting them.  Fortunately, there exists a {\em quantum swapping} process~\cite{zukowski1993event} as depicted in Fig.~\ref{fig:swap}: if nodes A and C share a Bell pair and nodes B and C share a separate Bell pair, then C can implement a {\em swap operation}, which we write as \iswap{C}{A}{B} whose steps bear resemblance to the teleportation process.  The swap ends with A and B sharing a Bell pair, with both of C's formerly-entangled qubits collapsed via measurement, ending their entanglement, with C's 2-bit measurement result sent to either A or B who can make the necessary local corrections to complete the swap.  In this paradigm, C is often referred to as a {\em repeater}, and, more generally, A and B can construct a shared Bell pair by performing a series of swaps across a {\em path} of repeaters~\cite{briegel1998quantum}, where each node $N_i$ along the path shares a Bell pair with the next hop, $N_{i+1}$, in the path.

The swapping process along a path shares many characteristics to routing in a traditional network setting where data flows along the path instead of entanglement swaps.  As in the traditional networking setting, there may exist multiple feasible paths over which swapping can connect an origin-destination pair, with differing paths having differing costs.  The costs of concern in the swapping regime are also familiar to the networking community: the number of swaps needed grows with the number of hops in the path, such that longer paths will likely take more time and require more repeater resources.  Also, differing paths may offer differing {\em fidelities}~\cite{jozsa1994fidelity} that provide a measure of the intactness of Bell Pairs, similar to how different classical networking paths can offer different reliabilities due to variations in corruption or packet loss along each path.  And, just as classical networks have a means to compensate for different reliabilities (e.g., error correction, selective retransit), quantum swapping also has a means to compensate for varying fidelities (a combination of quantum error correction (QEC) and a process known as {\em distillation} or {\em purification}~\cite{bennett1996purification}).

Not surprisingly, the swapping process has noticeable differences from classical network formulations.  The first is the magic of teleportation: that a complex quantum state can be moved by the combination of local operations and 2-bit classical transmission.  Second, the Bell pairs that are swapped carry no explicit payload and, aside from where the pair of qubits reside, are {\em interchangeable}: a Bell pair  used at any point in the swapping process to support a teleportation between nodes A and B could, if needed, be diverted to the swapping process to support teleportation between nodes C and D.  Thus, each Bell pair whose qubits reside at nodes $N_1$ and $N_2$ are effectively identical and for our purposes we refer to any such Bell pair as \share{$N_1,N_2$}.
In contrast, a data packet has an explicit payload which rarely can be diverted to support another application.  

\begin{figure}[h] 
    \centering
  
        \includegraphics[width=\linewidth]{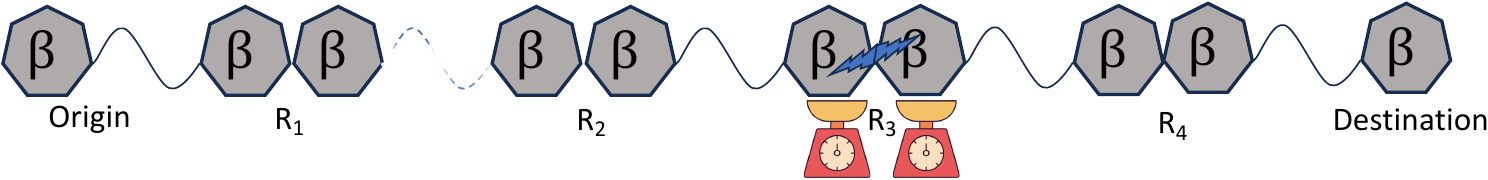}
        \caption{A swapping path example}
        \Description{A figure demonstrating a swapping process along a path}
        \label{fig:swap-path}

\end{figure}

The interchangeable nature of Bell pairs leads to a third noticeable distinction from traditional networking: unlike classical routing where each node performing its routing action must await the data item from the node one hop closer to the origin, the order in which repeaters perform their swapping action is arbitrary - any shuffle of the order in which a node on the path executes its swap will succeed, with the repeater performing the swap effectively extracting itself from the chain of entanglements that connects origin to destination: it replaces its two entanglements that connect it to its neighbors with a single entanglement directly entangling those neighbors.  This scenario is depicted in Fig.~\ref{fig:swap-path}, in which repeater $R_3$ performs \iswap{$R_3$}{$R_2$}{$R_4$}, collapsing a \share{$R_2$, $R_3$} and a \share{$R_3$,$R_4$} to form a \share{$R_2$,$R_4$}.  In our example, this swap is performed on the path even before $R_1$ and $R_2$ have yet to even establish a Bell pair: a phenomenon that cannot happen in classical communication.

\paragraph{Planned-path swapping:}
To our knowledge, all prior art that explores the distribution via swapping of Bell pairs through a quantum network uses what we refer to as a {\em planned-path} approach~\cite{pant2019routing,Halder2024OptimalRouting,chakraborty2019distributed,li2022connection,xiao2023connectionless}:  When a pair of nodes desires a Bell pair between them (i.e., for teleportation), a set (perhaps of size 1) of possible paths are identified, one is selected, and swaps are subsequently performed along this path.   Such approaches can be classified into {\em connection-oriented}~\cite{li2022connection} and {\em connectionless}~\cite{xiao2023connectionless}, where the former immediately maps Bell pairs to a specific swap path while the latter permits multiple paths that might both have interest in the same swap configuration (i.e., \iswap{C}{A}{B}) to compete over the Bell pairs that can implement this swap).  However, both approaches require the idea of identifying the path prior to implementing the swapping process.

Planned-path approaches overlook an important evolution that took place in the classical networking community for applications where there is a distributed demand for the same content~\cite{myers1999performance,stoica2003chord,ratnasamy2001scalable,schollmeier2001definition}.  Restricting communication along pre-defined paths can overlook potential efficiencies of scale, even when attempts are made to converge toward well performant paths~\cite{andersen2001resilient,chu2000case}.  In short, flexible strategies that pre-position content in a way that makes it more easily accessible when the demand arises or increases are preferred to only trying to finding an optimal path from static points at the immediate instance of demand.

In this paper, we consider the benefits and challenges toward constructing a quantum Internet using path-oblivious approaches where swapping is used to blaance Bell pairs holistically toward needed locations.  We begin in \S\ref{sec:tradeoffs} with early assumptions about cost that targeted early designs toward planned-path approaches, and why current evolution in the field supports the design of path-oblivious approaches.  In \S\ref{sec:LP}, we show how to formulate the path-oblivious approach as a linear program.  We present a simple distributed balancing algorithm in \S\ref{sec:algorithm}, compare its performance to planned-path approaches in \S\ref{sec:performance}, describe some preliminary thoughts on future directions in \S\ref{sec:layering}, and end with related work and conclusion respectively in \S\ref{sec:related} and \S\ref{sec:conclusion}.

\section{Planned v. Oblivious}
\label{sec:tradeoffs}

In this section, we justify why the quantum networking community fixates on planned-path approaches, and motivate why we think oblivious-path approaches have a future.

Planned-path approaches are (as we experienced in classical networking~\cite{zhang1993rsvp,gallassi1989atm}) initially appealing because if well organized, they can make optimal use of networking resources~\cite{breslau1998best}.  They are needed in high-cost, under-utilized or limited capacity networks.  But as the network scales and costs come down, the complexity of reserving access on multiple paths can bottleneck the supposed gain in efficiency.  So let us first present arguments why over the short term, planned-path approaches make sense.

\noindent {\bf Decoherence}: Even if left alone, Bell pairs naturally decohere, i.e., the entangled qubits naturally disentangle~\cite{briegel1998quantum}.  Planned-path advocates note that current and near-term coherence times are so short, that the entire swapping process must be completed in milliseconds before pairs to be used along the path collapse~\cite{victora2020purification}.  Applying a collapsed Bell pair anywhere in the swapping/teleportation process corrupts the measurement.  

\noindent {\bf Fidelity}: A further complication is the stochastic nature of quantum information.  Since measurement collapses the entanglement, it is impossible to validate via measurement that entangled quantum state is in its desired form.  Fidelity can be thought of as a confidence measure of how likely a given quantum state is in fact in its desired form~\cite{jozsa1994fidelity}.  One approach to increasing fidelity is a process called {\em distillation} a.k.a. {\em purification}~\cite{bennett1996purification}: a predictive process that uses (and destroys) one Bell pair to assess the correctness of another.    Because of decoherence, the fidelity of a  Bell pair must be continuously reaffirmed.  More recently, quantum error correction (QEC) approaches have been considered as an alternate and likely less costly approach to maintaining high fidelity~\cite{jiang2009quantum}, and there is ongoing research that considers hybrid approaches~\cite{pattison2024fast}.

\noindent {\bf Waste:} It should be clear that Bell pairs that are used for teleportation are in fact the product of the consumption of numerous Bell pairs that were used to form its entanglement (via swapping) and ensure high fidelity (via distillation).  It is wasteful to manufacture such a Bell pair  and not consume it before it decoheres, or to delay its use and unnecessarily incur additional costs of continual distillation or QEC.

\noindent {\bf Classical overheads:} Finally, swapping, teleportation, and distillation all require transmission of (albeit only a few bits) classical information across the network between points that wish to maintain entanglements.  Both planned-path and path-oblivious approaches will need to account for this classical transmission, as well as any additional classical coordination to (in the case of planned-path) establish and execute swapping routes, and (in the case of path-oblivious) learn about the status of the distributed state of Bell pairs such that a path-oblivious algorithm can make informed decisions about how to proceed.

\subsection{A Push Analogy: Water systems}
While in the near-term each Bell pair should be considered an expensive, short-lived entity, based on current progress trends~\cite{liu2024road}, we optimistically anticipate that as physicists and engineers put their heads together, the cost of generation will decline to a point where each Bell pair can simply be viewed as a drop of water in a very large bucket. This water analogy is intentional: path-planned approaches are analagous to a water park that only turns on the water for a ride when a person (valuable commodity) actively engages the ride and turns the water off immediately on completion.  Water, while not free, is in many places sufficiently prevalent such that approaches that ``push'' water to the faucet are preferable, even if water is wasted (e.g., pipes or taps drip, samples are used for testing, etc.)  As costs associated with Bell pairs drop, we speculate that a similar ``push'' architecture will in the long-term be preferable to path-based approaches, in that the small loss of efficiency will be offset by the significantly simpler control mechanism that loosens the need for planning and reservation.

Anticipating that over the next decades, Bell pair generation costs will drop and coherence times will grow, one can expect such networks to experience the same provisioning boom that classical networks have experienced over the last few decades.  This will presumably lead to a similar realization within the Quantum Internetworking community that scalability and flexibility should trump optimality.






\section{Path-oblivious LP formulation}
\label{sec:LP}

In contrast to a planned-path formulation that establishes paths for swapping to support a specific teleportation, a path-oblivious formulation can be thought of in its crudest form as ``swap gossip'': a repeater that holds two Bell pairs, each entangled with a different node, can implement a swap to directly entangle those neighbors without necessarily supporting any explicit teleportation goal: it is simply trying to help ``spread'' the availability of different pairs of nodes connected by Bell pairs.  If this spreading is well-designed, then as teleportation needs appear, the Bell pairs supporting them may already be present, or may simply be a small number of swaps away.

We can best understand the asymptotic capabilities of a path-oblivious approach by formulating the generation and consumption of entangled state as a linear flow program.  Our formulation presumes an ideal environment where each repeater has a limitless buffer to store its qubit components of Bell pairs, and generated Bell pairs  do not decohere.  

The problem is formulated over a set of nodes $\setofnodes$.  Each pair of nodes $x,y \in \setofnodes$ has an average rate $g(x,y) = g(y,x) \ge 0$ at which they can, from scratch, {\em generate} Bell pairs \share{$x,y$}, where some (perhaps many) pairs of $x,y$ are unable to generate pairwise, such that $g(x,y) = 0$ (e.g., perhaps they are physically too far from one another to form Bell pairs without the assistance of swapping).  Similarly there are pairs $x,y$ that {\em consume} Bell pairs \share{$x,y$} (i.e., to implement teleportation) at an average rate of $c(x,y) \ge 0$, where many pairs may have no need for consumption, such that $c(x,y) = 0$.  We can define the {\em generation graph}, $G$,  consisting of undirected edges $(x,y)$  for which $g(x,y)>0$.  Observing that a swap never increases the number of Bell pairs held at a node, any node $x$ that wishes to be part of the quantum Internet must satisfy $g(x,y) > 0$ for some $y$, and that $\sum_y c(x,y) \le \sum_y g(x,y)$ if it is to maintain its consumption rate.  Further, if a generation graph $G$ is not fully connected, then it is impossible to build (via generation and swaps) a Bell pair that entangles $x$ with $y$ when they lie in distinct connected components.  Last, we are only interested in the generation and consumption of Bell pairs that are distributed between nodes, such that we assume $g(x,x) = c(x,x) = 0$.

We can observe the asymptotic rate at which node $i$ implements the swap \iswap{$i$}{$x$}{$y$} and write this rate as $\swap_i(x,y) = \swap_i(y,x)$ where $\swap_i(x,i) = \swap_i(x,x) = 0$ since these result in Bell pairs entangled at a single node.

\subsection{Steady State Analysis}

For each pair $x,y$, we denote $\rateup(x,y)=\rateup(y,x)$ to be the asymptotic rate at which entanglements arrive between $x$ and $y$ and $\ratedn(x,y)=\ratedn(y,x)$ the rate at which they depart:
\setlength{\abovedisplayskip}{5pt}
\setlength{\belowdisplayskip}{5pt}
\setlength{\abovedisplayshortskip}{0pt}
\setlength{\belowdisplayshortskip}{3pt}
\begin{align}
    \rateup(x,y) &=& g(x,y) + \sum_i \swap_i(x,y)\\
    \ratedn(x,y) &=& c(x,y) + \sum_i (\swap_x(i,y) + \swap_y(i,x))
\end{align}
i.e., a Bell pair arrives between $x$ and $y$ when it is explicitly generated or when some third node $i$ implements \iswap{$i$}{$x$}{$y$}, and a Bell-pair departs between $x$ and $y$ when it is explicitly consumed or when either \iswap{$x$}{$i$}{$y$} or \iswap{$y$}{$i$}{$x$}.

Since Bell pairs cannot depart at a rate faster than they arrive, we must always have $\ratedn(x,y) \le \rateup(x,y)$, and in steady-state, we expect $\ratedn(x,y) = \rateup(x,y)$ for all pairs $x,y$.

\subsection{Decoherence, Distillation, and QEC}

Our simple LP can easily be extended to handle constant rates of decoherence and distillation per node-pair.  We assume that the pair $x$ and $y$ only utilize Bell pairs for consumption and swapping after their fidelity has achieved a sufficient level which is accomplished with an expected number $\distill_{x,y}$ of distillations, and that fully distilled Bell pairs are discarded (lost) at a rate of $\decohererate_{x,y}$, we can reformulate $\rateup$ and $\ratedn$ as
\begin{align}
\rateup(x,y) = \decohererate_{x,y} (g(x,y) + \sum_i \swap_i(x,y))\\
 \ratedn(x,y) = \distill_{x,y} (c(x,y) + \sum_i (\swap_x(i,y) + \swap_y(i,x)))
 \end{align}

 Alternatively, QEC can be added simply by assuming that QEC is applied to generated Bell pairs, after which QEC is the form of correction.  If the overhead of the QEC (i.e., the number of physical qubits per logical qubit) is ${\mathcal R}$, we can simply thin the generation rate by $g(x,y)$ to be $g(x,y) / \mathcal{R}$

 \subsection{Optimization Objectives}

It is worth noting that in our formulation, one can think of the external inputs being the $\{g(x,y)\}$ and  $\{c(x,y)\}$ which are respectively determined by the physical architecture's capabilities and teleportation demand, with the $\{\swap_i(x,y)\}$ being the set of variables that need to be solved.  The optimal setting of the $\{\swap_i(x,y)\}$ depend of course on the desired optimization criterion.  We list a few possibilities:

\begin{fitemize}

\item When generation is sufficient to handle consumption, then conserve generation: either minimize $\sum_{x<y} g(x,y)$ or alternatively, minimize the maximum $g(x,y)$.

\item When generation is insufficient to handle consumption, then reduce consumption fairly.  More formally, let $\gamma_{x,y}$ $\kappa_{x,y}$ respectively be the maximum possible generation rate and desired consumption rate for pair $x,y$.  Then, in addition to solving for $\{\swap_i(x,y)\}$, also find $\{g(x,y)\}$ and $\{c(x,y)\}$ such that $g(x,y) \le \gamma_{x,y}$ and $c(x,y) \le \kappa_{x,y}$ where one maximizes $\sum_{x<y} c(x,y)$, or alternatively maximizes the minimum $c(x,y) < \kappa(x,y)$, or alternatively finds the largest $\alpha$ such that $c(x,y) = \alpha \kappa_{x,y}$.

\item If we first maximize the consumption objective, and subsequently reformulate and minimize generation rate with respect to the redefined consumption rates, the generation rate must be the optimal, which implies that when distillation processes are being run, they are being performed in the most efficient manner.

\end{fitemize}

\section{Max-min distributed swapping}
\label{sec:algorithm}

While our LP formulation assists with understanding asymptotic behavior of a path-oblivious swapping system, it does not address the practicalities of how and when to swap given the stochastic nature of generation, consumption, distillation and decoherence processes.  We present a simple {\em balancing} approach that, as an initial baseline, appears to work well.

We suppose for now that each node $x$ maintains a count $\bcount_x(y)$ of the number of Bell pairs it stores that are shared with each $y$ in the network (note $\bcount_x(y) = \bcount_y(x)$).  For any two neighbors $y$ and $y'$, we say that the swap \iswap{$x$}{$y'$}{$y$} is {\em preferable} when $\bcount_y(y') + 1 \le \min(\bcount_x(y)-\distill_{x,y}, \bcount_x(y')-\distill_{x,y'})$, such that $x$ would only reduce its own counts if doing so aids a pair whose count would be no larger after the swap.  If $x$ has more than one preferable swap candidate, it performs a swap with minimal $\bcount_y(y')$.

Were generation and consumption to both cease, this swapping process would balance bell pairs in the network such that the minimum count would be maximized, i.e., when no swap candidates are preferrable, a max-min fair allocation has been established, in that no buffer count can be increased without reducing another that was already smaller~\cite{jaffe1981bottleneck}.  Intuitively, as Bell pairs are generated, swaps will flow from the points of generation, and as Bell pairs are consumed or used in swaps, swaps will flow toward these points of drainage.

Note that in this initial formulation of the algorithm, we are not concerning ourselves with classical costs, which would include not only the 2 bits required to complete the swap, but also the sharing of information between all nodes of the $\numnodes \choose 2$ edge counts.  This may very well be feasible given existing and emerging high speed classical networks.  Still, we discuss possible ways to reduce classical communication overheads in \S\ref{sec:layering}.

\section{Performance Analysis}
\label{sec:performance}

To evaluate the efficacy of our baseline algorithm, we simulate a Quantum Internet on a variety of topologies, where a subset of repeater pairs (that form a connected graph)  act as generators and a (possibly overlapping) subset of repeater pairs act as consumers.  The generators generate Bell pairs and these Bell pairs are swapped using the baseline algorithm.  Consumers can then consume these Bell pairs.

We consider two generation graph topologies in this paper: a {\em cycle graph} containing $\numnodes$ numbered $0$ through $\numnodes-1$ where  $g(x,y)>0 \iff y=x \pm 1 \mod \numnodes$, and an embedding on a wraparound $\sqrt{\numnodes} \times \sqrt{\numnodes}$ grid.  Each node $x$'s position can be described by coordinates $(x_i,x_j), 0 \le i,j < \sqrt{\numnodes}$ where $g(x,y) >0$ can only occur when either $x_i = y_i$ and $x_j = y_j \pm 1 \mod \sqrt{\numnodes}$ or $x_j=y_j$ and $x_i = y_i \pm 1 \mod \sqrt{\numnodes}$.  Generation edges are added uniformly at random on the grid until the underlying generation graph connects all nodes.  In these preliminary simulations, when $g(x,y) = 1$ for all edges where $g(x,y)>0$.   

We generate 35 pairs from the set of all possible $\numnodes \choose 2$ pairs as our consumers, and construct a sequence of consumption requests from these pairs that must be satisfied in the order of the sequence.\footnote{This was done to prevent biasing the cost toward easy-to-satisfy pair requests - a point addressed in \S\ref{sec:layering}.}  All nodes perform the swapping process at an identical rate.  We found that varying this rate did not signficantly alter the results presented below.

To measure the efficiency of our swapping process, we compare the number of swaps our algorithm performs in comparison to the minimum number of swaps needed were each consumption event satisfied by swaps along the shortest path in the underlying generation graph.  Specifically, if the consumption pair is $(x,y)$, we identify the shortest path and compute the number of swaps needed to form the Bell pair that entangles $x$ and $y$.  When the length of the shortest path between $x$ and $y$ is $n$ hops and all distillation overheads $\distill_{x,y}$ are set to an identical value of $\distill$, the optimal swapping process (known as nested swapping) requires $s(n)$ swaps where $s(1) = 0, s(2) = \distill$ and $s(n) = \distill (s(\lfloor n/2 \rfloor) + s(\lceil n/2 \rceil))$ for $n>2$.  The {\em swap overhead} of our distributed algorithm is defined the number of swaps performed in simulation divided by $\sum_c s(\ell(c))$ where the sum over $c$ covers all consumption events that were satisfied in simulation, and $\ell(c)$ is the number of hops in the shortest path within the generation graph for consumption event $c$.  Note this measure can be no less than 1, since the denominator is the minimum number of swaps that could satisfy all consumption events.  However, there are several reasons why this scoring is conservative, in that our algorithm probably does better against practical planned-path approaches:


\begin{fitemize}
    \item Practical planned-path approaches need not always take the shortest swapping path (i.e., in attempts to load balance or avoid congestion).  For similar reasons, the optimal nested swapping order that minimizes distillation costs may not always be practical.
    \item When the simulation ends, there are multiple swaps that occurred within the simulation that were not used for consumption (i.e., they would still be available for future consumptions).
\end{fitemize}

\begin{figure}[ht] 
    \centering
  
        \includegraphics[width=0.9\linewidth]{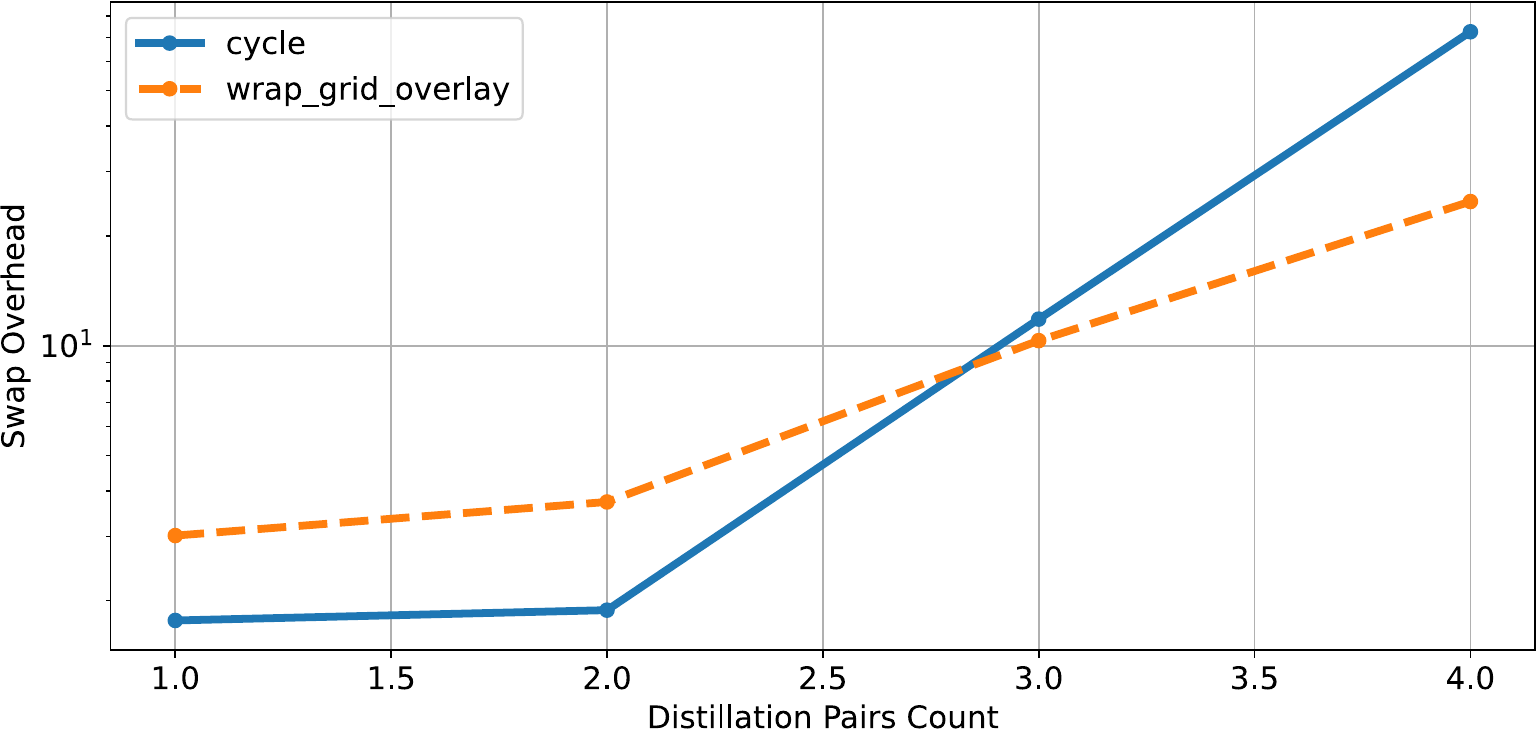}
        \caption{Simulation results $\numnodes=25$, varying $\distill$}
        \Description{Simulation results varying distillation}
        \label{fig:vary-distill}
   
\end{figure}

\begin{figure}[ht] 
    \centering
  
        \includegraphics[width=0.9\linewidth]{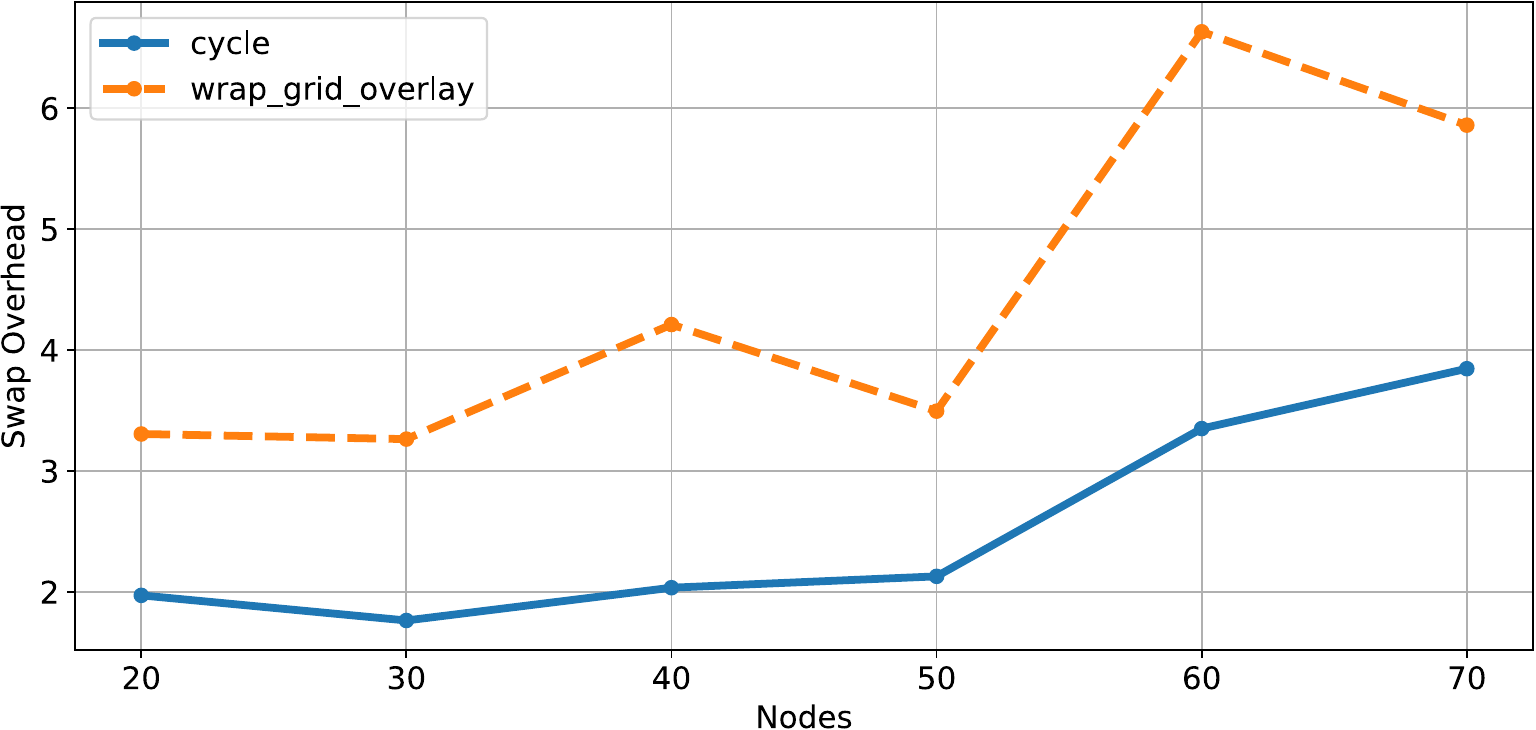}
        \caption{Simulation results, $\distill=1$, varying $\numnodes$}
        \Description{Simulation results varying $\numnodes$}
        \label{fig:vary-numnodes}
   
\end{figure}

Figures  \ref{fig:vary-distill} and \ref{fig:vary-numnodes} plot the swap overhead on the $y$-axis as $\distill$ and $\numnodes$ are respectively varied along the $x$-axis with $\numnodes = 25$ and $\distill=1$ respectively in thre graphs.  Our preliminary results show that the overhead is expected to grow slowly as the number of nodes in the graph is increased.  In contrast, the overhead grows exponentially as $\distill$ is increased.  This is likely due to the swapping process straying from the most distillation-efficient nested swapping ordering.

In summary, as quantum memories, coherence times, and fidelity preserving mechanisms improve, even a simple baseline approach is reasonably efficient in comparison to more stringent planned-path approaches, and there are plenty of dimensions over which path-oblivious protocols can be further tuned.

\section{Discussion}
\label{sec:layering}

\noindent {\bf Baseline} Our initial attempt at a path-oblivious swapping algorithm makes no attempts to optimize the problem beyond a simple max-min balancing scheme, and how best to refine the path-oblivious swapping algorithm to further reduce overhead is an ongoing trajectory of our research.  If costs in high-distillation regimes seem excessively high, there are several things to do.  For instance, reducing the likelihood that node $i$, very distant from both $x$ and $y$ in the generation graph, implements a swap between $x$ and $y$ when they are close to one another in the generation graph. 

\noindent {\bf Realistic coherence, QEC and distillation overheads} Admittedly, our models of coherence time and distillation processes are oversimplified, and our use has been simply to ballpark how these additional phenomena might affect performance of path-oblivious approaches.  Future study should better incorporate the actual effects of such processes, and consider protocol adjustments that might result (e.g., avoiding combining Bell pairs with short expected remaining coherence times with those that have longer times).

\noindent{ \bf Classical overheads} If classical communication overheads are of concern, the immediate global knowledge of all buffers can likely be relaxed as well.  For instance, a BitTorrent-like approach~\cite{cohen2003incentives} with a similar choke/unchoke mechanism, where each node knows only the status of a rotating but small number of neighbors, would intuitively scale well.

\noindent{\bf Toward a layered Quantum Internet Stack} Path-oblivious approaches also give rise to a cleaner separation within a layered approach.  For instance, the max-min algorithm we describe can be thought of as a network-layer protocol, whereas decisions about distillation processes and rejection of aged Bell pairs can be thought of as a transport layer functionality that ``cleanses'' the Bell pairs passed through the system.

\noindent{\bf Hybrid oblivious with minimal planning} Path-oblivious can also be viewed as a "seeding" for requests.  If the Bell pair is not immediately available upon consumption request, the consuming pair can then find a shortest path among the existing Bell pairs (which could be much shorter than their shortest path on the underlying graph).  Doing this could help mitigate the starvation effect we saw in our simulations, where consumption requests between nodes who are close on the generation graph would usurp the Bell pairs needed to form the longer paths.

\section{Related Work}
\label{sec:related}

Prior work in quantum network resource allocation often assumes short coherence times, motivating reactive, planned-path approaches that generate Bell pairs on-demand and rely on explicit path selection \cite{pant2019routing,Halder2024OptimalRouting,chakraborty2019distributed,li2022connection} with some such as \cite{xiao2023connectionless} proposing a connectionless variant where the path is reserved but Bell pairs are not, such that criss-crossing paths compete for available Bell pairs at shared links. Our LP formulation is similar in flavor to that of~\cite{chakraborty2020entanglement}.  However, that formulation is significantly more complex in that it attempts to identify specific paths that prioritize high fidelity that neglects application of distillation and quantum error correction processes.

Proactive strategies~\cite{chakraborty2019distributed} propose pre-preparing virtual links but assumes short lifetimes, requiring frequent regeneration. \cite{talsma2024continuously} studies continuous distribution in regular topologies under fidelity decay but offers no optimization framework or support for distillation. \cite{inesta2023optimal} analyzes cutoffs in chain topologies and references purification, though operational integration remains limited.
Percolation-based studies such as \cite{acin2007entanglement} explore probabilistic global connectivity but do not account for distillation, memory dynamics, or network-wide optimization. Surveys and architecture proposals  \cite{Abane2025EntanglementRouting,bacciottini2025leveraging,sun2024message} offer taxonomies or protocol-level mechanisms but do not model fidelity-aware resource distribution.

In contrast, our work assumes long-lived quantum memories (motivated by advances such as \cite{Reagor2016QuantumMemory,lei2023quantum}) and explicitly incorporates distillation costs. We present both a centralized linear program and a distributed protocol for proactive, path-oblivious, and connectionless Bell pair distribution. This enables efficient, fidelity-aware resource balancing across arbitrary topologies, supporting asynchronous many-to-many traffic without relying on precomputed paths.

\section{Conclusion}
\label{sec:conclusion}

As coherence times increase, swapping mechanisms will become less dependent on real-time construction of the end-to-end path, such that we anticipate approaches migrating away from path-based methods to more share-oriented methods.  The interchangeability of Bell pairs makes the formulation of quantum internetworking extremely amenable to path oblivious methods that distribute the availability of such state in a manner that is divorced from the underlying topological constraints.  While the interplay between distillation and QEC remains an important issue to resolve, the classical networking community's experience with massive distribution of content emphasizing stateless approaches will become more relevant and the classical networking community's expertise could have positive impact on this important genre.  For instance, considering how to layer a transport layer atop such a novel networking layer.

\bibliographystyle{plain}  
\bibliography{refs}

\end{document}